\documentclass{article} % For LaTeX2e
\usepackage{nips13submit_e,times}
\usepackage{url}
\usepackage{graphicx}
\usepackage{amsmath}
\usepackage{amssymb}

\usepackage{booktabs}  % professionally typeset tables
\usepackage{textcomp}  % better copyright sign, among other things
\usepackage{xcolor}
\usepackage{lipsum}    % filler text
\usepackage{subfig}    % composite figures
\usepackage{verbatim}

\title{Bayesian Neural Networks for Genetic Association Studies of Complex Disease}

\author{
Andrew L. Beam\thanks{ Corresponding author. }\\
Bioinformatics Research Center\\
North Carolina State University\\
Raleigh, NC\\
\texttt{albeam@ncsu.edu} \\
\And
Alison Motsinger-Reif \\
Bioinformatics Research Center\\
Department of Statistics \\
North Carolina State University\\
Raleigh, NC\\
\texttt{alison\_motsinger@ncsu.edu} \\
\And
Jon Doyle \\
Department of Computer Science \\
North Carolina State University\\
Raleigh, NC\\
\texttt{Jon\_Doyle@ncsu.edu} \\
}

\nipsfinalcopy % Uncomment for camera-ready version

\begin{document}

\maketitle

\begin{abstract}
\subsection*{Background}
Discovering causal genetic variants from large genetic association studies poses many difficult challenges. Assessing which genetic markers are involved in determining trait status is a computationally demanding task, especially in the presence of gene-gene interactions.
\subsection*{Results}
A non-parametric Bayesian approach in the form of a Bayesian neural network is proposed for use in analyzing genetic association studies. Demonstrations on synthetic and real data reveal they are able to efficiently and accurately determine which variants are involved in determining case-control status. Using graphics processing units (GPUs) the time needed to build these models is decreased by several orders of magnitude. In comparison with commonly used approaches for detecting interactions, Bayesian neural networks perform very well across a broad spectrum of possible genetic relationships.
\subsection*{Conclusions}
The proposed framework is shown to be powerful at detecting causal SNPs while having the computational efficiency needed to handle large datasets.

\end{abstract}

\section{Background}
The ability to rapidly collect and genotype large numbers of genetic variants has outpaced the ability to interpret such data, leaving the genetic etiology for many diseases incomplete. The presence of gene-gene interactions, or epistasis, is believed to be a critical piece of this “missing heritability” \cite{manolio:find}. This has in turn spurred development on advanced computational approaches to account for these interactions, with varying degrees of success \cite{motsinger:comparisonanalytical, motsinger:comparisonapproaches, koo:review}. The main computational challenge comes from the vast number of markers that are present in a typical association study. This problem is exacerbated when interactions between two or more markers must be considered. For example, given an experiment that genotypes 1,000 markers, examining all possible interactions between two of the markers involves consideration of nearly half a million combinations. This situation becomes exponentially worse as higher order interactions are considered. Modern genome-wide association studies (GWASs) routinely consider 1-2 million single nucleotide polymorphisms (SNPs), which would require examining half a trillion potential interactions. As whole genome sequencing (WGS) methods become commonplace, methods that cannot cope with large data sets will be of little utility. Data on this scale will require approaches that can find interactions without having to enumerate all possible combinations. As genotypic technology advances, datasets now routinely include millions of SNPs.

Several distinct types of methods have emerged that attempt to address this challenge. Perhaps one of the most popular approaches from the last decade has been Multifactor Dimensionality Reduction (MDR) \cite{moore:flexible,hahn:multifactor}, and extensions of the method. MDR is a combinatorial search that considers all possible interactions of a given order and selects the best model via cross validation. Because MDR is an exhaustive search, it suffers from the previously discussed scalability issue, though recent work using graphics processing units has attempted to lessen this deficit \cite{greene:multifactor}. MDR is reliant upon a permutation testing strategy to assess statistical significance for each marker, so the computational burden becomes prohibitive for large datasets. Permutation testing computes a p-value for-a statistic of interest (such as an accuracy measure from MDR) by randomly permuting the class labels and calculating the statistic on the permuted dataset. This procedure is repeated many times to compute a “null” distribution for the statistic of interest. The relative percentage of instances in the permuted null distribution that are less than or equal to the actual statistic from the unpermuted data is taken as the desired one-sided p-value. Unfortunately, this can be extremely expensive for large datasets when many hypotheses are simultaneously tested, leading to a large multiple testing scenario. To get the required resolution for a Bonferroni corrected p-value of 0.05 when considering a set of 1,000 SNPs, one must perform 20,000 permutations. This makes permutation testings infeasible for even moderately sized datasets. 

Another popular approach is Bayesian Epistasis Association Mapping (BEAM) \cite{zhang:bayesian}. BEAM partitions markers into groups representing individual (i.e. marginal) genetic effects, interactions, and a third group resenting background markers that are uninvolved with the trait. BEAM employs a stochastic Markov Chain Monte Carlo (MCMC) search technique to probabilistically assign markers to each group and uses a novel “B-statistic” based on the MCMC simulation to assign statistical significance to each marker. This allows BEAM to assign statistical significance without the need to perform a costly permutation test.  This method has been demonstrated successfully on data sets with half a million markers. However, the recommended amount of MCMC iterations needed is quadratic in the number of SNPs considered \cite{zhang:bayesian}, possibly limiting its effectiveness for larger datasets.

Many popular machine learning algorithms have also been adopted for use in analyzing association studies. Notable examples are decision trees (both bagged, i.e. random forests,  \cite{lunetta:screening,diaz:gene} and boosted \cite{li:detecting} support vector machines (SVM) \cite{guyon:gene}, Bayesian networks \cite{jiang:learning}, and neural networks \cite{motsinger:comparisonanalytical}. In particular, tree-based methods such as random forests and boosted decision trees have been found to perform well in several previous association studies \cite{lunetta:screening,li:detecting,diaz:gene}. Machine learning approaches are appealing because they assume very little \emph{a priori} about the relationship between genotype and phenotype, with most methods being flexible enough to model complex relationships accurately. However, this generality is something of a double-edged sword as many machine learning algorithms function as black boxes, providing investigators with little information on which variables may be most important. Typically it is the goal of an association study to determine which variables are most important, so a black box may be of little use. Some approaches have easy adaptations that allow them to provide such measures. Both types of tree based methods (bagged and boosted) can provide measures of relative variable importance \cite{breiman:random,friedman:greedy}, but these indicators lack measures of uncertainty, so they are unable to determine how likely a variable's importance measure is to occur by chance without resorting to permutation testing.

In this study, we propose the use of Bayesian neural networks (BNNs) for association studies to directly address some the issues with current epistasis modeling. While BNNs have been previously developed and applied for other tasks \cite{lisboa:bayesian,baesens:bayesian,neal:bayesian,neal:bayesiantraining} they have yet to see significant usage in bioinformatics and computational biology. Like most complex Bayesian models, BNNs require stochastic sampling techniques that draw samples from the posterior distribution, because direct or deterministic calculation of the posterior distribution is often intractable. These posterior samples are then used to make inferences about the parameters of the model or used to make predictions for new data. Standard MCMC methods that employ a random walk such as the Metropolis-Hastings (RW-MH) algorithm \cite{metropolis:equation, hastings1970monte} (which is the algorithm that forms the core of BEAM \cite{zhang:bayesian}) explores the posterior distribution very slowly when the number of predictors is large. If $d$ is the number of parameters in a model, the number of iterations needed to obtain a nearly independent sample is $O(d^2)$ \cite{neal2011mcmc} for RW-MH. This makes the RW-MH algorithm unsuitable for neural network models in high-dimensions, so the Hamiltonian Monte Carlo (HMC) algorithm is instead used to generate samples from the posterior. HMC has more favorable scaling properties, as the number of iterations needed is only $O(d^{5/4})$ \cite{neal2011mcmc}. HMC achieves this favorable scaling by using information about the gradient of the log-posterior distribution to guide the simulation to regions of high posterior probability. Readers familiar with standard neural network models will notice an inherent similarity between Bayesian neural networks sampled using HMC and traditional feed-forward neural networks that are trained using the well known back-propagation algorithm \cite{rumelhart:learning}, as both take steps in the steepest direction using gradient based information. Though HMC will in general explore the posterior distribution in a more efficient manner than RW-MH, the evaluation of the gradient can very expensive for large data sets. Recent work has shown that this drawback can be lessened through the use of parallel computing techniques \cite{beam:fast}. 

The BNN framework outlined here has several features designed to address many of the challenges inherent in analyzing large datasets from genetic association studies. These advantages are outlined below.
\begin{itemize}
\item Quantification of variable influence with uncertainty measures. This allows variable influence to be assessed relative to a null or background model using a novel Bayesian testing framework. This avoids reliance on a permutation testing strategy.
\item Automatic modeling of arbitrarily complex genetic relationships. Interactions are accounted for without having to examine all possible combinations. This is achieved from the underlying neural network model.
\item An efficient sampling algorithm. HMC scales much better than other MCMC methods, such as the RW-MH algorithm, in high-dimensions.
\item Computational expediency through the use of GPUs.  The time needed to build the model is greatly reduced using the massive parallel processing offered by GPUs.
\end{itemize}
We offer evidence for these claims using several simulated scenarios and a demonstration on a real GWAS dataset. In addition, we compare the proposed approach to several popular methods so that relative performance can be assessed. 

\section{Methods}
Neural networks are a set of popular methods in machine learning that have enjoyed a flurry of renewed activity spurred on by advances in training so-called “deep” networks \cite{hinton:improving,bengio2009learning}. The term neural network can refer to a very large class of modeling techniques, but to be clear we use the term here to refer to multilayer feed-forward perceptions (MLPs). In the most basic sense, neural nets represent a class of non-parametric methods for regression and classification. They are non-parametric in the sense that they are capable of modeling any smooth function on a compact domain to an arbitrary degree of precision without the need to specify the exact relationship between input and output. This is often succinctly stated as “neural nets are \emph{universal function approximators}“ \cite{hornik:multilayer}. This property makes them appealing for many tasks, including modeling the relationship between genotype and phenotype, because a sufficiently complex network will be capable of automatically representing the underlying function. 

Some draw backs of classical neural nets are natural consequences of their strengths. 
Due to their flexibility, neural nets are highly prone to “over-fitting” to data used to train them. Over-fitting occurs when the network starts to represent the training data exactly, including noise that may be present, which reduces its ability to generalize to new data. Many methods exist to address this issue, but popular methods such as weight decay are well known to be approximations to a fully Bayesian procedure \cite{neal:bayesian, williams:bayesian}. Another issue with standard neural nets is they are often regarded as “black boxes” in that they do not provide much information beyond a predicted value for each input. Little intuition or knowledge can be gleaned as to which inputs are most important in determining the response, so nothing coherent can be said as to what drives the network’s predictions. Discussions of the advantages and disadvantages of neural nets for gene mapping have been reviewed in \cite{motsinger:neural}. First we describe the base neural network model, and then describe how this can be incorporated into a Bayesian framework. 

The network is defined in terms of a directed, acyclic graph (DAG) where inputs are feed into a layer of hidden units. The output of the hidden units are then fed in turn to the output layer which transforms a linear combination of the hidden unit outputs into a probability of class membership. Specifically consider a network with \emph{p} input variables, \emph{h} hidden units, and 2 output units to be used to predict whether an observation belongs to class 1 or class 2. Let $x_i=<x_{i1},…,x_{ip}>^T$ be the input vector of $p$ variables and $y_i=<y_{i1},y_{i2}>$ be the response vector, where $y_{i1} = 1$ if observation i belongs to class 1 and 0 if not, with $y_{i2}$ is defined in the same way for class 2. Hidden unit $k$ first takes a linear combination of each input vector followed by a nonlinear transformation, using the following form:

\begin{equation}\label{hiddenunit}
h_k(x_i) = \phi \left(b_k + \sum_{j=1}^p w_{kj}*x_{ij} \right)
\end{equation}
where $\phi(\cdot)$ is a nonlinear function. For the purposes of this study, consider the logistic transformation, given as:

\[
\phi(z) = \frac{1}{1+e^{-z}}
\label{logistic}
\]
Several other activation functions such as the hyperbolic tangent, linear rectifier, and the radial basis/Gaussian functions are often used in practice. Each output unit takes a linear combination of each $h_k$ followed by another nonlinear transformation. Let $f_1(x_i)$ be the output unit that is associated with class 1:

\begin{equation} \label{outputunit}
f_1(x_i) = \psi\left(B_1 + \sum_{k=1}^h W_{k1}*h_k(x_i)\right)
\end{equation}
Note we have used upper case letters to denote parameters in the output layer and lowercase letters to indicate parameters belonging to the hidden layer. The  $\psi(\cdot)$ function is the softmax transformation of element $z_1$ from the vector $z = <z_1,\dots,z_n>$:
\[
\psi(z_1) = \frac{\exp(z_1)}{\sum \limits_{i=1}^n\exp(z_i)}
\]
In this representation, $f_1 (x_i )$ represents the estimated conditional probability that $y_i$ belongs to class 1, given the input vector $x_i$. A similar definition is made for output unit 2, $f_2 (x_i )$. Note that for the case of only 2 classes, $f_2(x_i)=1-f_1(x_i)$ because the softmax transformation forces the outputs to sum to 1.

Having described the formulation for standard neural networks we next describe how this can be extended using the Bayesian formulation. Bayesian methods define a probability distribution over possible parameter values, and thus over possible neural networks. To simplify notation, let $\theta=\{B,W, \beta,w\}$ represent all of the network weights and biases shown in equations \ref{hiddenunit}, \ref{outputunit}. The posterior distribution for $\theta$ given the data $x_i$ and $y_i$, is given according to Bayes’ rule:

\begin{equation} \label{bayesrule}
p(\theta | x_i,y_i) = \frac{L(\theta|x_i,y_i)\cdot \pi(\theta)}{m(x)}
\end{equation}
where $m(x) = \int L(\theta|x_i,y_i) \cdot \pi(\theta) d\theta$ is the marinal density of the data. $L(\theta|,x_i,y_i)$ is the \emph{likelihood} of $\theta$ given then data and $\pi(\theta)$ is the prior distribution for the network parameters. However, in practice we only need to be able to evaluate the numerator of (\ref{bayesrule}) up to a constant because we will be relying on MCMC sampling techniques that draw from the correct posterior without having to evaluate $m(x)$, which may be intractable in high dimensions. 

Often it is better to work with the log-likelihood $l(\theta|x_i,y_i )=\log⁡(L(\theta|x_i,y_i ))$, because the raw likelihood can suffer from numerical overflow or underflow for large problems. In this study we assume the log-likelihood for a neural network with 2 output units is binomial:
\begin{align} \label{loglike}
l(\theta | x_i,y_i )=y_{i1} \cdot \log⁡(f_1 (x_i))+y_{i2} \cdot \log⁡(1-f_1 (x_i))
\end{align}
Next every parameter in the model must be given a prior distribution. The prior distribution codifies beliefs about the values each parameter is likely to take before seeing the data. This type of formulation is extremely useful in high-dimensional settings such as genomics, because it enables statements such as `most of the variables are likely to be unrelated to this response' to be quantified and incorporated into the prior. In this study, we adopt a specific prior structure known as the \emph{Automatic Relevance Determination} (ARD) prior. This prior was originally introduced in some of the foundational work on Bayesian neural nets \cite{neal:bayesian,neal:assessing} and later used in SVMs for cancer classification \cite{li2002bayesian}. 

The ARD prior groups network weights in the hidden layer together in a meaningful and interpretable way. All of the weights in the hidden layer that are associated with the same input variable are considered part of the same group. Each weight in a group is given a normal prior distribution with mean zero and a common variance parameter. This shared group-level variance parameter controls how large the weights in a group are allowed to become and performs shrinkage by pooling information from several hidden units, which helps to prevent overfitting \cite{neal:assessing}.  Each of the group-level variance parameters is itself given a prior distribution, typically an Inverse-Gamma distribution with some shape parameter $\alpha_0$ and some scale parameter $\beta_0$. These parameters often referred to as hyper parameters, and can themselves be subject to another level of prior distributions, but for the purposes of this study, we will leave them fixed as user specified values. Specifically, for a network with $h$ hidden units, the weights in the hidden layer for input variable $j$ will have the following prior specification:

\[
w_{j1},\ldots,w_{jp}\sim N(0,\sigma_j^2)
\]
\[
\sigma_j^2 \sim IG(\alpha_0,\beta_0)
\]
This structure allows the network to automatically determine which of the inputs are most relevant. If variable $j$ is not very useful in determining whether an observation is a case or a control, then the posterior distribution for $\sigma_j^2$ will be concentrated around small values. Likewise, if variable $j$ is useful in determining the response status, most of the posterior mass for $\sigma_j^2$ will be centered on larger values. 

\subsection{Hamiltonian Monte Carlo (HMC) for Neural Networks}
Here we briefly give an overview of Hamiltonian Monte Carlo (HMC) for neural networks, but please see \cite{neal:bayesian,neal:bayesiantraining, neal:mcmc} for a thorough treatment. HMC is one of many Markov Chain Monte Carlo (MCMC) methods used to draw samples from probability distributions that may not have analytic closed forms. HMC is well suited for high-dimensional models, such as neural nets, because it uses information about the gradient of the log-posterior to guide the sampler to regions of high posterior probability. For neural networks, we adopt the two-phase sampling scheme of Neal used in \cite{neal:bayesian}. In the first phase, we update the values of the variance parameters using a Gibbs update, conditional on the current values of the network’s weights. In the next phase, we leave the variance parameters fixed and update the network weights using HMC. We repeat this procedure of Gibbs-coupled HMC updates until we have acquired the desired number of posterior samples.

The higher level variance parameters, including those in the ARD prior, have simple closed forms because the Inverse-Gamma distribution is a conditionally conjugate prior distribution for the variance parameter of a Normal distribution. To obtain a new value for a variance parameter, conditional on the values of the weights a parameter controls, one makes a draw from the following Inverse-Gamma distribution:
\begin{align} \label{gibbs}
\sigma^2_{new} \sim IG \left(\alpha_0 + \frac{n_w}{2} , \beta_0 + \sum_{i=1}^{n_w} \frac{w_i^2}{2} \right)
\end{align} 
where each $w_i$ is a weight controlled by this variance parameter, $n_w$ is the number of weights in the group, and $\alpha_0$, $\beta_0$ are the shape and scale parameters respectively of the prior distribution.				

HMC then proceeds by performing L number of “leap-frog” updates for the weights, given the values of the variance parameters. The algorithm introduces a fictious momentum variable for every parameter in the network that will be updated by simulating Hamiltonian dynamics on the surface of the log-posterior. Since HMC was orginally created in statistical physics it is often presented in terms of “energy potentials” which is equivalent to an exponentiation of the negative log-posterior, but we will describe the algorithm directly in terms of the log-posterior, which is more natural for our purposes. The full algorithm for sampling the posterior of all parameters (network weights and variance parameters) is shown below.
\begin{figure}[!h]
\centering
\includegraphics[scale=0.65]{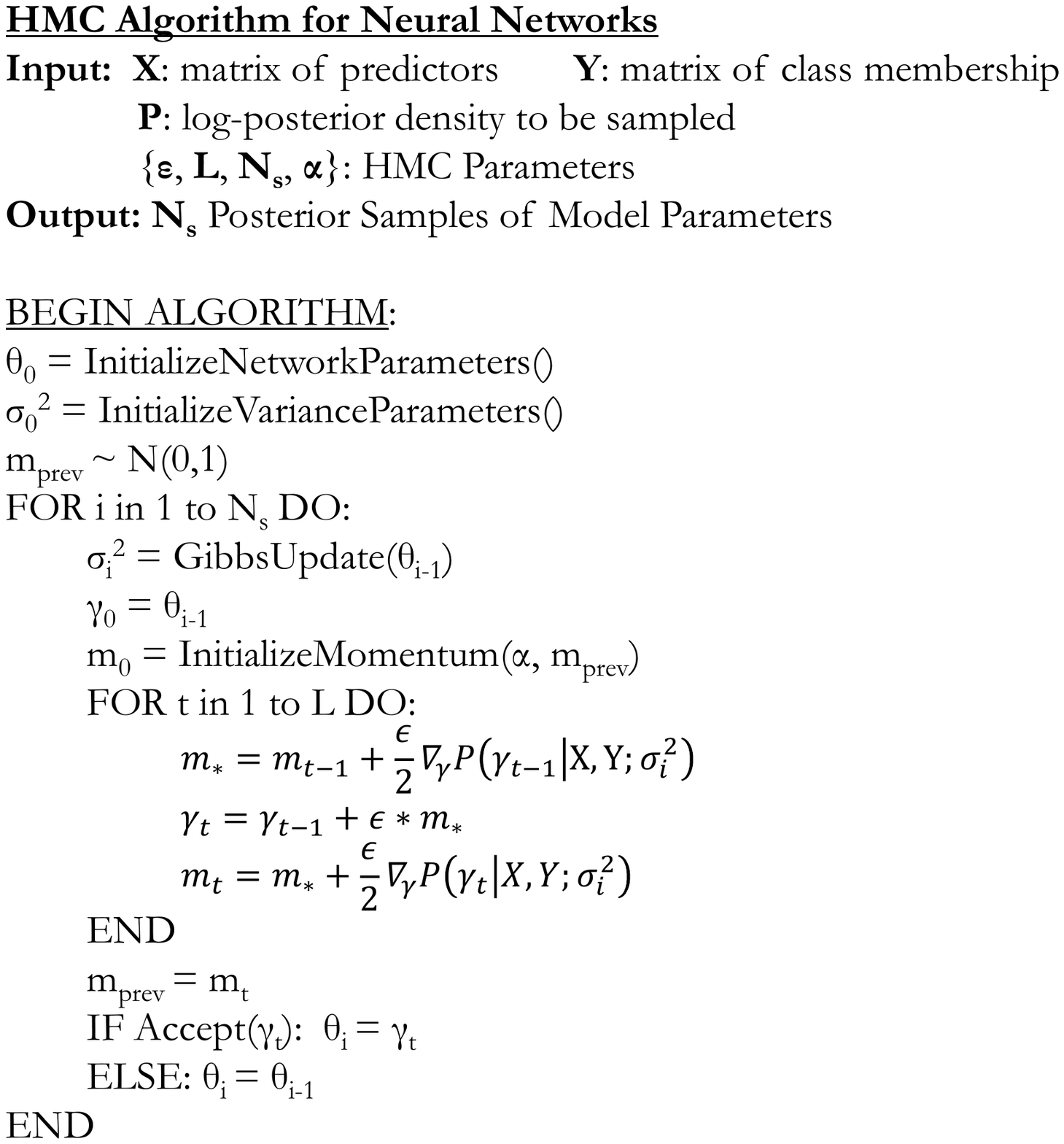}
\caption{Algorithm for HMC-based posterior sampling for the neural network model.}
\label{fig:hmcbnn}
\end{figure}

A few details in the HMC algorithm as shown need further explanation. First the momentum variables (m) are refreshed after every sequence of L leap-frog updates, shown in the algorithm as \texttt{InializeMomentum($\alpha$)}. In the most simple formulation each momentum component is a independent draw from a Normal distribution, with mean 0 and standard deviation of 1. However, this can lead to wasted computation because the sampler may start out in bad direction by chance, requiring more leap-frop updates until the sampler is heading in a useful direction resulting in random-walk like behavior. To combat this, we use the persistent momentum refreshes \cite{nabney:netlab, neal:bayesian} which initializes the momentum using a weighted combination of the final momentum value of the previous leap-frog update and a draw of standard normal random variable. Using the notation of the algorithm, this is shown below:
\begin{align*}
m_0 = \alpha*m_{prev}+ \sqrt{1-\alpha^2 }*\zeta
\end{align*}
where $\zeta \sim N(0,1)$. If the proposal is rejected the momentum must be negated. This must be done to ensure that the canonical distribution is left intact \cite{nabney:netlab}. This formulation reduces the number of leap-frog updates (L) needed to reach a distant point by suppressing random-walk behavior while leaving the correct target distribution of the Markov chain intact. \cite{nabney:netlab,neal:bayesian,neal:mcmc}. The \texttt{Accept}$(\gamma_t)$ function returns true if the new proposal, $\gamma_t$, is accepted according to a modified Metropolis-Hastings acceptance probability. \texttt{Accept}$(\gamma_t)$ returns true with the following probability:
\begin{align*}
\bar{\alpha} = \min\left(1,\frac{P(\gamma_t|X,Y;\sigma_i^2) - \frac{1}{2}m_t^Tm_t}{P(\gamma_{t-1}|X,Y;\sigma_i^2) - \frac{1}{2}m_{t-1}^Tm_{t-1}}\right)
\end{align*}
However, the posterior distribution is often very `bumpy' with many posterior modes \cite{neal:bayesian}. This property may be exacerbated in high dimensions, so becoming stuck in one mode for extended periods of time is a great concern. To alleviate this we modify the acceptance probability, $\bar{\alpha}$, to correspond to a \emph{flattened} version of the posterior whose acceptance probability is given as $\alpha^* = \bar{\alpha} \cdot T$, for $ T > 1$, which is equivalent to sampling from $P(\theta|X,Y;\sigma_t^2)^{\frac{1}{T}}$. While its true that we are no longer sampling from the exact posterior $P(\theta|X,Y;\sigma_t^2)$, under mild regularity conditions the posterior modes of the correct distribution remain intact \cite{andrieu2003introduction}. Since none of the parameters have biological interpretations modifying the posterior in this way of little concern, if the full procedure is capable of maintaining a  favorable discriminatory ability. We find the trade-off between ease of sampling across a wide-range possible scenarios and exactness of the posterior to be acceptable.

\subsection{HMC Using Graphics Processing Units (GPUs)}
Previous work has shown how the gradient and log-posterior evaluations needed by HMC can be sped-up by as much as 150x for large problems using Graphics Processing Units (GPUs) \cite{beam:fast}. We adopt that framework here and express the gradient calculations as matrix-vector operations or element-wise operations. Similarly, evaluation of the log-posterior can be expressed in terms of linear operators and element-wise operations. Using GPUs for these operations is well known in the neural network literature \cite{bergstra2010theano,noel:gpu,kyoung:gpu} as the gradient of the log-posterior corresponds roughly to the well-known back-propagation algorithm and evaluation of the log-posterior corresponds to the “feed-forward” operation in standard neural networks. However, to our knowledge this study represents the first GPU-enabled implementation of Bayesian neural networks. Without GPU computing, it is likely that the computational burden imposed by large datasets would be too great for the Bayesian neural network framework to be feasible. 

All of methods discussed in this study are implemented in the Python programming language. All GPU operations were conducted using the Nvidia CUDA-GPU \cite{nvidia2008programming} programming environment and accessed from Python using the PyCuda library \cite{klockner2012pycuda}. Source code containing the Bayesian neural network package is available at \url{https://github.com/beamandrew/BNN}.

\subsection{Bayesian Test of Significance for ARD Parameters}
Given the ARD prior a natural question to ask is how large do values of $\sigma_j^2$ need to be for input j to be considered relevant compared to a variable that is completely unrelated. This question can be framed in terms of a Bayesian hypothesis test. In this framework we will assume that under the “null hypothesis” a variable is completely irrelevant in determining the status of the response. If this were a simple linear model, this would be equivalent to saying the regression coefficient for this variable has a posterior mean of zero. In the neural network model the ARD parameters that determine how relevant each input is are strictly positive, so we need a baseline or null model for the ARD parameters in order to determine if we can “reject” this null hypothesis of irrelevance. In order to construct and test this hypothesis, we make a simplifying assumption that weights for unrelated variables have a normal distribution with mean 0 and variance $\sigma_{null}^2$ i.e. $w_{kj} \sim N(0,\sigma_{null}^2)$. Due to the complex statistical model, the true posterior distributions for the weights under the null may not be exactly normal, but this approximation will be useful in simplifying the calculations. Additionally since, the prior for each weight is normal, this approximation will most likely not be too far from true posterior form under the stated null hypothesis.

Since $\sigma_{null}^2$ represents the “null” ARD parameter associated with a variable of no effect, we wish to test whether a variable of interest is “significantly” greater than this null value. We use the phrases “null” and “significance” here because of their familiar statistical connotations, but they should not be confused with the p-value based frequentist hypothesis testing procedure, as we are operating within a fully Bayesian framework. Our goal becomes testing whether the mean, $\mu_j$ of the posterior distribution for the ARD parameter, $\sigma_j^2$,  is greater than the mean of the null, $\mu_{null}$, for the null ARD parameter $\sigma_{null}^2$. Specifically we wish to test the following null hypothesis:

\begin{align*}
H_0: \mu_{null}= \mu_j
\end{align*}
against the one-sided alternative:
\begin{align*}
H_a: \mu_{null} < \mu_j
\end{align*}
To test this, we need to know the closed form of $\mu_{null}$. Making use of the iterative two-stage sampling scheme, we will derive this form by induction. We will also make use of several well-known facts of random variables. Firstly, if a random variable $X$ has an inverse-gamma distribution, i.e. $X \sim IG(\alpha,\beta)$, then the mean or expected value of $X$, $E[X]$, is given by $\frac{\beta}{\alpha-1}$. Next, if the sequence of random variables $X_1 \ldots X_n$ are each independently and identically distributed as $N(0,\sigma^2)$, then $\sum \limits_{i=1}^n \left(\frac{X_i}{\sigma} \right)^2 = <\frac{X_1}{\sigma},\ldots,\frac{X_n}{\sigma}>^T<\frac{X_1}{\sigma},\ldots,\frac{X_n}{\sigma}> \sim \chi^2_n$, i.e. a chi-squared random variable with $n$ degrees of freedom. This sum has an expected value of n, from the definition of a chi-squared random variable. This implies the conditional expected value $E[<X_1,…,X_n >^T<X_1,…,X_n>|\sigma]=\sigma^2*n$. Using these basic facts we will show that under the null, the two-stage sampling scheme leaves expected value of the ARD parameter invariant, i.e. $\mu_{null}=\mu_{prior}$.

For a network with $h$ hidden units, let $w_j= <w_{1j},\ldots ,w_{hj}>$ be a vector containing all of the weights associated with input $j$, where each component of $w_j$ is initially distributed according to the prior, $N(0,\sigma_j^2 )$ and $\sigma_j^2 \sim IG(\alpha_0,\beta_0 )$. The mean, for $\sigma_j$ at the start of the simulation is $\mu_0=\frac{\beta_0}{\alpha_0 -1}$. We begin the simulation at iteration i=1 and perform a Gibbs update of $\sigma_j^2$. The Gibbs update for the shape parameter, $\alpha_1=\alpha_0+n_w/2$, is iteration independent and will remain fixed for the entirety of the simulation. However, the Gibbs update for the scale parameter, $\beta_1= \beta_0+(w_j^T w_j)/2$, depends upon the current values of the weights, and thus will take on a random value at each iteration. However, we can compute the expected value for $\beta_1$ as:
\begin{align*}
E[\beta_1]&= E\left[\beta_0+\frac{w_j^T w_j}{2}\right]\\
&=\beta_0+ \frac{1}{2} E[w_j^T w_j ]\\
&=\beta_0+\frac{1}{2} (h \cdot E[\sigma_0^2 ])\\
&=\beta_0+\frac{h}{2} \cdot \frac{\beta_0}{\alpha_0-1}\\
&= \beta_0+\frac{h}{2} \cdot\mu_0\\
\end{align*}
Thus, the expected value of $\beta_1$ after the first Gibbs update is $\beta_0+ \frac{h}{2} \cdot \mu_0$. Note that this expectation is independent of simulation iteration, so this result will hold for all $\beta_1,\beta_2,\cdots,\beta_𝑡$. Next, we use this fact to compute the expected value of the ARD parameter, $E[\sigma_1^2]$:
\begin{align*}
E[\sigma^2_1] &= E \left[ \frac{\beta_1}{\alpha_0 + h/2 - 1} \right]\\
&= \frac{E[\beta_1]}{\alpha_0 + h/2 - 1}\\
&= \frac{\beta_0+h/2 \cdot \mu_0}{\alpha_0+h/2-1}\\
&= \frac{\beta_0}{\alpha_0-1+h/2}+ \frac{h/2 \cdot \mu_0}{\alpha_0-1+h/2}\\
&= \frac{\frac{1}{\alpha_0 -1}}{\frac{1}{\alpha_0 -1}} \cdot \frac{\beta_0}{\alpha_0-1+h/2}+ \frac{h/2 \cdot \mu_0}{\alpha_0-1+h/2}\\
&=  \frac{\mu_0}{1+h/2 \cdot \frac{1}{\alpha_0-1 }}+ \frac{h/2 \cdot \mu_0}{\alpha_0-1+h/2}\\
&= \mu_0 \left( \frac{1}{1+h/2 \cdot \frac{1}{\alpha_0-1}}+ \frac{h/2}{\alpha_0-1+h/2} \right) \\ 
&= \mu_0 \left( \frac{\alpha_0-1}{\alpha_0-1+h/2}+ \frac{h/2}{\alpha_0-1+h/2} \right) \\
&= \mu_0 \left( \frac{\alpha_0-1+h/2}{\alpha_0-1+h/2} \right) \\ 
&= \mu_0\\
\end{align*}
Thus, the Gibbs update of the ARD parameter does not change the expected value under the null, since we defined $E[\sigma_0^2]=\mu_0$. This establishes the base case, and now we show the induction step. Given $E[\beta_{t+1}]=E[\beta_t]= \beta_0+\frac{h}{2} \cdot \mu_0$ and $E[\sigma_t]= \mu_0$ then:

\begin{align*}
E[\sigma_{t+1}] &= E\left[ \frac{\beta_{t+1}}{\alpha_0 + h/2 -1} \right]\\
&= \frac{E[\beta_{t+1}]}{\alpha_0 + h/2 -1} \\
&= \frac{\beta_0 + h/2 \cdot \mu_0}{\alpha_0 + h/2 -1} \\
&= \mu_0
\end{align*}
where the simplification between lines 3 and 4 proceeds as before. This concludes the proof. 

\section{Results}
\subsection{Existing Methods Used for Comparison}
We selected several methods to serve as baselines for evaluation of the BNN’s performance. As previously mentioned BEAM and MDR are widely used methods and so were included in our evaluation. We used a custom compiled 64-bit version of BEAM using the source provided on the website \cite{zhang:website} of the authors of \cite{zhang:bayesian}.  The java-based MDR package was downloaded from the MDR source-forge repository (http://sourceforge.net/projects/mdr/) and called from within a Python script. To evaluate the effectiveness of tree-based methods, we used an approach nearly identical to that in \cite{li:detecting}, which was based on boosted decision trees. The boosted decision tree model provides measures of relative influence for each variable that indicate how important a given variable is, relative to the others in the model. To fit the boosted tree model we used the \texttt{gbm} package in R. Finally, we also included the standard 2 degrees-of-freedom chi-square test of marginal effects.

As discussed, some approaches such as MDR and GBM require a permutation testing strategy to assess statistical significance. This makes assessing their performance on large datasets difficult, due to the amount time required to perform the permutation test. During our pilot investigations on a dataset containing 1,000 SNPs, each individual run of MDR was found to take roughly 1 minute to complete. The time needed to complete the required 20,000 permutations would be roughly 2 weeks. If we wish to evaluate a method’s effectiveness on hundreds or thousands of such datasets, this run time becomes prohibitive. As such, we divided our primary analysis into two sections. In the first section, we evaluated methods that \emph{do not} rely on permutation testing on datasets containing 1,000 SNPs each. However, since we wish to compare the results of the BNN to that of MDR and GBM, we performed a second set of analyses on smaller datasets that only contained 50 SNPs each, for which permutation testing is feasible.  This two-pronged strategy allowed us to evaluate a wide range of popular approaches in a reasonable amount of time, while serving to underscore the need for methods that do not rely on permutation testing.

\subsection{Parametric Models of Multi-Locus Relationships}
In this section we performed an analysis of three biallelic models of genotypic relationships. These models have been used previously \cite{zhang:bayesian, li:detecting} and are meant to reflect theoretical and empirical evidence for genetic relationships involving multiple loci \cite{li:complete}. Tables \ref{tab:additive}, \ref{tab:threshold}, and \ref{tab:epistatic} contain the risk relative of disease for each genotype combination, where a capital and lower case letters represent the major and minor alleles, respectively.
\begin{table}[!h]
     \caption{Additive Risk Model}
 \label{tab:additive}
  \label{tab:add}
  \begin{center}
    \begin{tabular}{cccc}
      \toprule
      Genotype & AA & Aa & aa \\
      BB & $\eta$ & $\eta(1+\theta)$ &$\eta(1+2\theta)$ \\
      Bb & $\eta(1+\theta)$ & $\eta(1+2\theta)$ & $\eta(1+3\theta)$ \\
      bb &  $\eta(1+2\theta)$ & $\eta(1+3\theta)$ & $\eta(1+4\theta)$ \\
      \bottomrule
    \end{tabular}
  \end{center}
\end{table}
\begin{table}[!h]
     \caption{Threshold Risk Model}
 \label{tab:threshold}
  \label{tab:thresh}
  \begin{center}
    \begin{tabular}{cccc}
      \toprule
      Genotype & AA & Aa & aa \\
      BB & $\eta$ & $\eta$ &$\eta$ \\
      Bb & $\eta$ & $\eta(1+\theta)$ & $\eta(1+\theta)$ \\
      bb &  $\eta$ & $\eta(1+\theta)$ & $\eta(1+\theta)$ \\
      \bottomrule
    \end{tabular}
  \end{center}
\end{table}

\begin{table}[!h]
     \caption{Epistatic Risk Model}
 \label{tab:epistatic}
  \label{tab:epi}
  \begin{center}
    \begin{tabular}{cccc}
      \toprule
      Genotype & AA & Aa & aa \\
      BB & $\eta$ & $\eta$ &$\eta(1+4\theta)$ \\
      Bb & $\eta$ & $\eta(1+2\theta)$ & $\eta$ \\
      bb &  $\eta(1+4\theta)$ & $\eta$ & $\eta$ \\
      \bottomrule
    \end{tabular}
  \end{center}
\end{table}

The symbols $\eta$ and $\theta$ in the tables represent the baseline risk and effect size, respectively. We simulated genotypes for the disease SNPs for a range of minor allele frequencies (MAFs) and simulated the disease status for 1,000 cases and 1,000 controls using the risks given in Tables \ref{tab:additive}, \ref{tab:threshold}, and \ref{tab:epistatic}. We embedded the causal SNPs in a background of 998 non-causal SNPs, for a total of 1,000 SNPs to be considered. For each combination of effect size, $\theta \in \{0.5,1.0,1.5,2.0\}$, MAF$\in \{0.1,0.2 ,0.3,0.4,0.5\}$, and model type (Additive, Threshold and Epistasis) we generated 100 datasets. This yielded a total of 6,000 datasets for evaluation. All datasets in this section were created using the R statistical programming language \cite{r:language}. 

We ran BNN, BEAM, and the $\chi^2$ test on each dataset and recorded whether or not both disease SNPs were declared as significant by each method. We took the fraction of datasets where both disease SNPs were correctly identified as an estimate of statistical power. For BEAM and the $\chi^2$ test, we used the canonical Bonferroni corrected significance threshold of $p < 0.05$. We used the recommended parameter settings for BEAM \cite{zhang:bayesian} and performed $1*10^6$ sampling iterations for each dataset. For the BNN approach, we used a network with 1 hidden layer and 5 logistic units and a softmax output layer with 2 units. The network parameters in the hidden layer are given ARD priors, while the network parameters in the output are given a common Gaussian prior.  The hyper parameters for the Inverse-Gamma prior for the ARD parameters were $\alpha_0=5$,$\beta_0=2$ while the hyper parameters for the Gaussian priors were $\alpha_0=0.1$,$\beta_0=0.1$. The parameters for the HMC algorithm were $\epsilon=5*10^{-2}$, $L=15$, $\alpha=0.75$, and $T=5*10^3$. The cutoff value for the novel Bayesian ARD testing framework was 0.6. We discarded the first 25 samples as burn-in and kept 100 samples to be used for inference. Processing of each dataset by the BNN took approximately 3 minutes. The results are shown below in Figures \ref{fig:add}, \ref{fig:thresh} and \ref{fig:epi}.

%\begin{lscape}
\begin{figure}[!h]
\centering
\includegraphics[width=\textwidth]{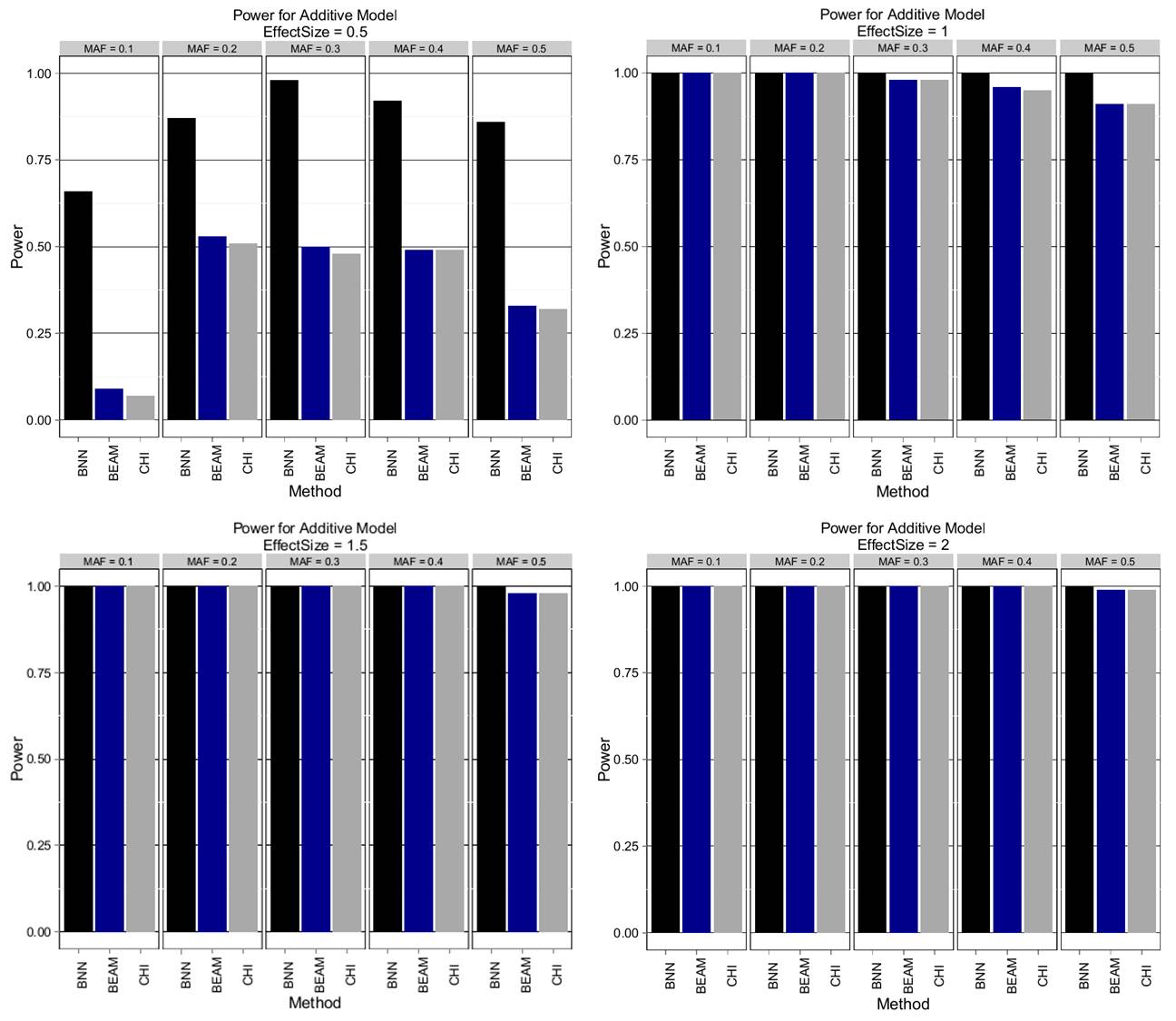}
\caption{Additive Model. Estimated power to detect both disease SNPs using Bayesian neural networks (BNN), BEAM, and $\chi^2$ test (CHI) with 2 d.f. Effect sizes of $\{0.5, 1.0, 1.5, 2.0\}$ are shown in order from left to right, top to bottom. Within each pane results are stratified by minor allele frequency (MAF).}
\label{fig:add}
\end{figure}
%\end{lscape}

%\begin{lscape}
\begin{figure}[!h]
\centering
\includegraphics[width=\textwidth]{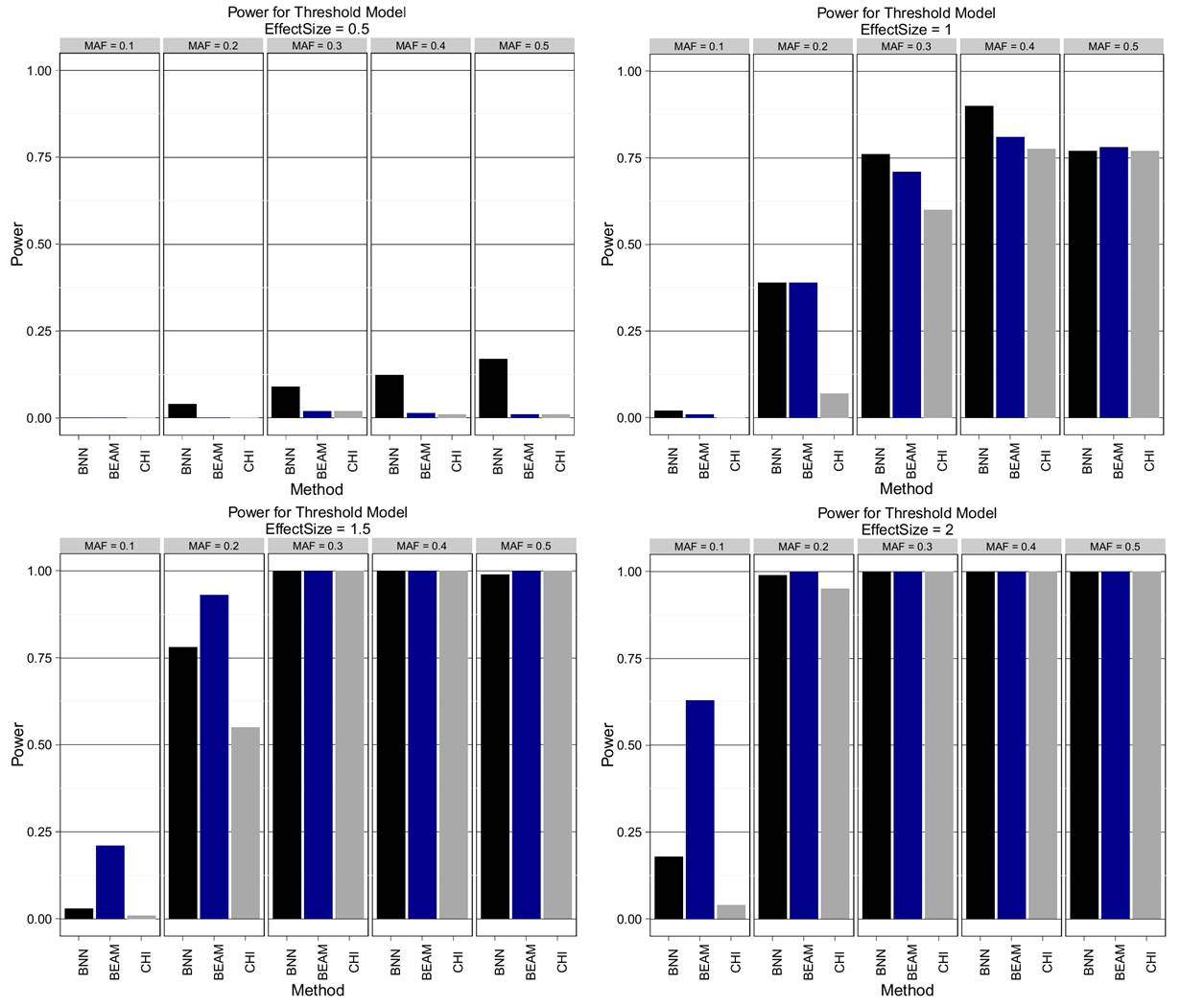}
\caption{Threshold Model. Estimated power to detect both disease SNPs using Bayesian neural networks (BNN), BEAM, and $\chi^2$ test (CHI) with 2 d.f. Effect sizes of $\{0.5, 1.0, 1.5, 2.0\}$ are shown in order from left to right, top to bottom. Within each pane results are stratified by minor allele frequency (MAF).}
\label{fig:thresh}
\end{figure}
%\end{lscape}

%\begin{lscape}
\begin{figure}[!h]
\centering
\includegraphics[width=\textwidth]{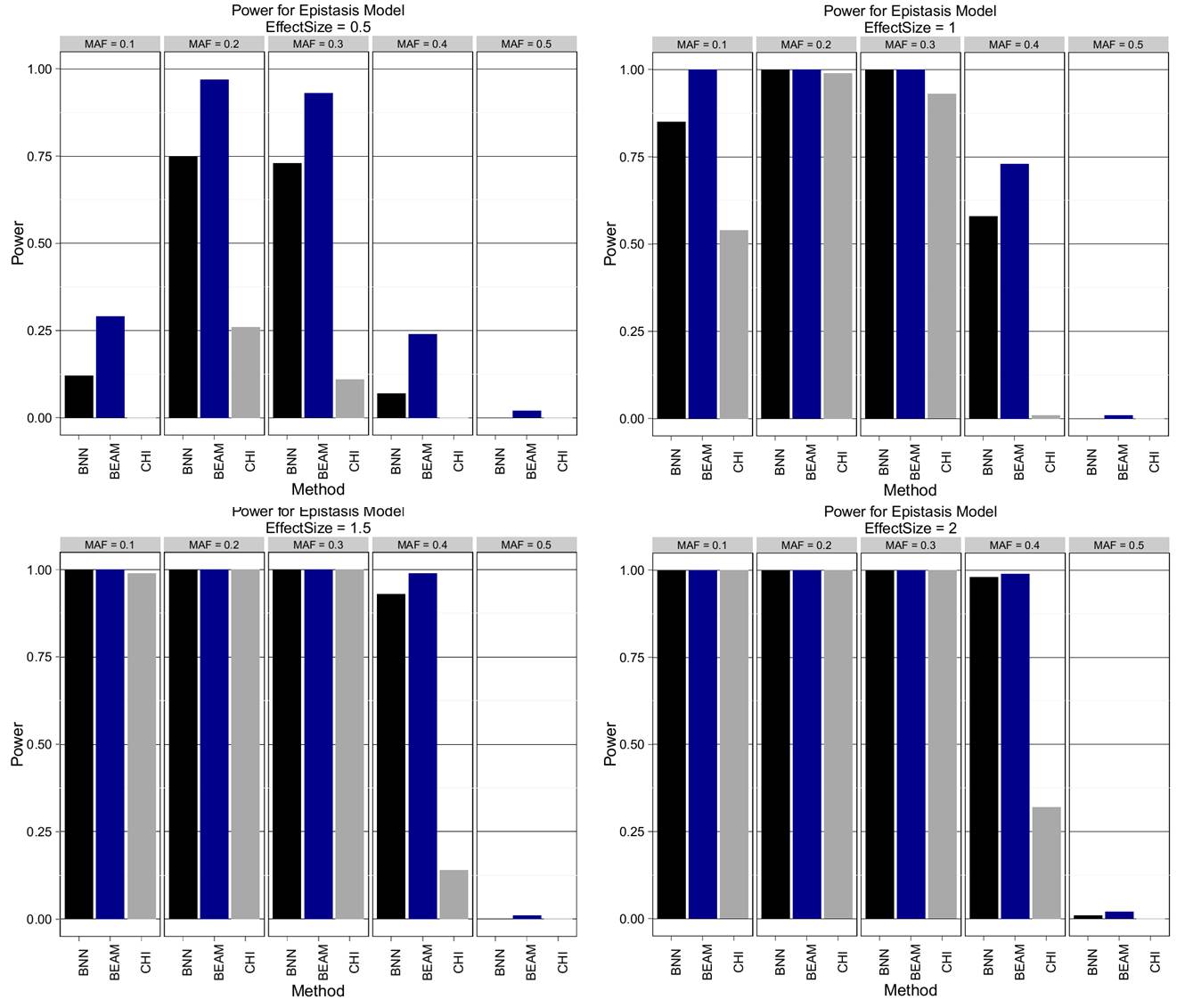}
\caption{Epistatic Model. Estimated power to detect both disease SNPs using Bayesian neural networks (BNN), BEAM, and $\chi^2$ test (CHI) with 2 d.f. Effect sizes of $\{0.5, 1.0, 1.5, 2.0\}$ are shown in order from left to right, top to bottom. Within each pane results are stratified by minor allele frequency (MAF).}
\label{fig:epi}
\end{figure}
%\end{lscape}

BNNs were found to be uniformly more powerful than both BEAM and the $\chi^2$ test for the additive model. BNNs show excellent power, even for small effect sizes and achieve 100\% power for second smallest effect size across all tested MAFs. In contrast, BEAM showed relatively little power for the smallest effect size and never achieves 100\% for all MAFs, even at the highest level of effect size. The threshold model tells a similar story. For all but 3 combinations of MAF and effect size, the BNN model is again uniformly more powerful than both BEAM and the χ2 test. The picture from the epistatic model is slightly more mixed. BEAM appeared to do a better job at the smallest effect size, while performing equally well as BNNs on the remaining three effect size levels. All three methods had almost no power to detect the causal SNPs for a MAF of 0.5. These results suggest that BNN is uniformly more powerful the $\chi^2$ test for these genetic models, and may be more powerful than BEAM in most instances.

\subsection{Simulated Epistatic Relationships without Marginal Effects}
In this section, we evaluated the performance of all the methods examined in the previous section (BNN, BEAM, and the $\chi^2$) as well as GBM and MDR. Since MDR and GBM rely on permutation testing, we reduced the size of the dataset to accommodate this strategy. To generate test datasets, we used the GAMETES software package \cite{urbanowicz:gametes}. This package allows users to specify the proportion of variance for case/control status that is due to genetic variants (i.e. broad-sense heritability) as well as how many loci are involved in determining trait status. These relationships are generated such that there are minimal marginal effects, resulting in relationships that are nearly purely epistatic. Relationships without marginal effects are in some sense `harder' than those with marginal effects, because the causal SNPs contribute to trait status only through their interaction. Preliminary analysis on the reduced SNP datasets indicated that if the same models were used as in the previous section, most methods would have nearly 100\% power for all simulated scenarios, which would provide little useful feedback for discerning which approaches were working best. This was the primary motivation for using the `harder', purely epistatic relationships instead of the parametric models we used previously.

Using GAMETES, we analyzed two levels of heritability (5\% and 10\%) across a range of MAFs (0.05, 0.1, 0.2, 0.3, 0.4, 0.5). Power was measured as in the previous section using 100 instances for each heritability/MAF combination for a total of 1200 data sets used in evaluation. The results are shown below in Figures \ref{fig:5nomargin} and \ref{fig:10nomargin}.

%\begin{lscape}
\begin{figure}[!h]
\centering
\includegraphics[width=\textwidth]{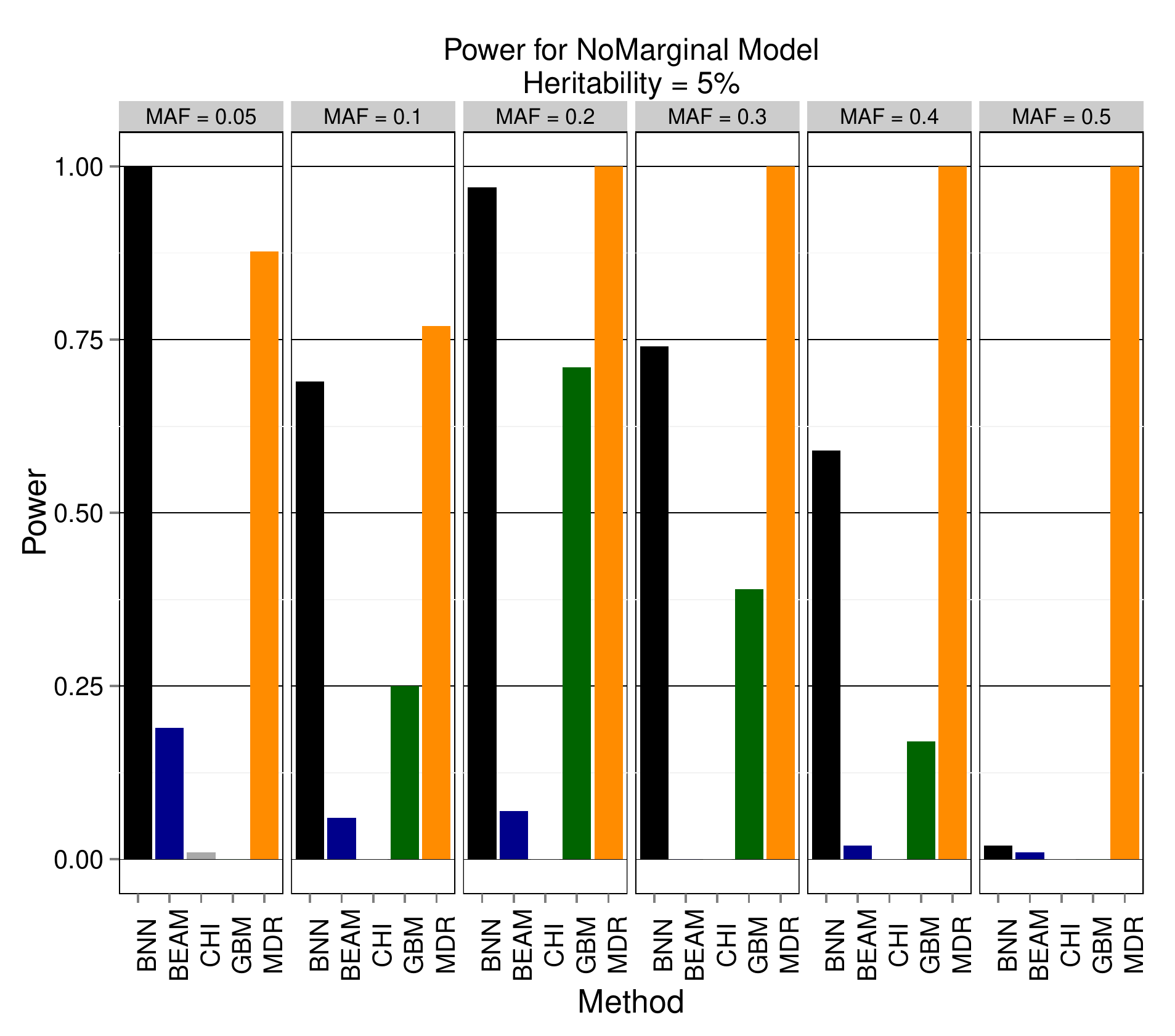}
\caption{Purely Epistatic Model with 5\% heritability. Estimated power to detect both disease SNPs of Bayesian neural networks (BNN), BEAM, χ2 test (CHI) with 2 d.f., gradient boosted trees (GBM), and MDR.  The results are stratified by minor allele frequency (MAF).}
\label{fig:5nomargin}
\end{figure}
%\end{lscape}

%\begin{lscape}
\begin{figure}[!h]
\centering
\includegraphics[width=\textwidth]{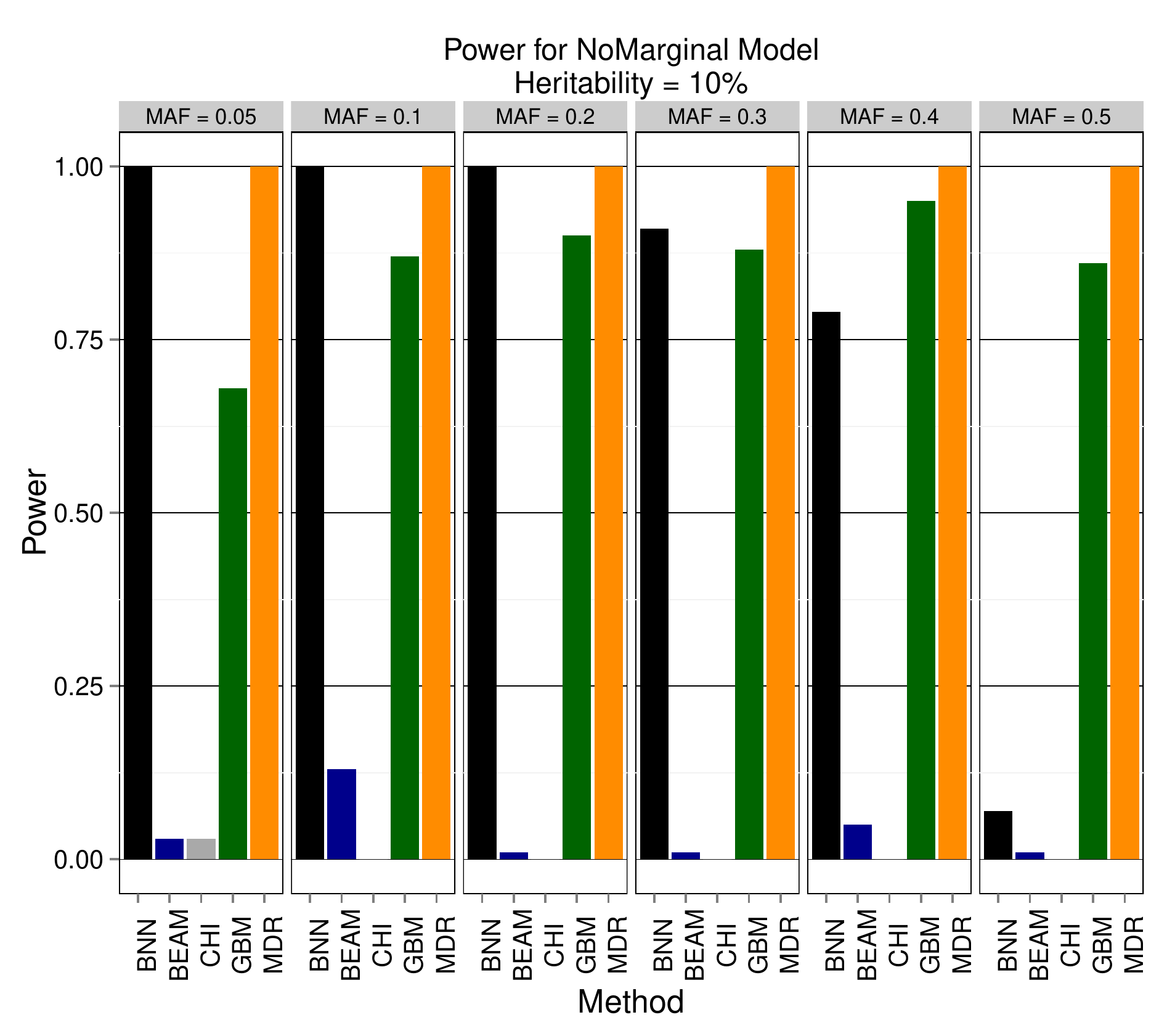}
\caption{Purely Epistatic Model with 10\% heritability. Estimated power to detect both disease SNPs of Bayesian neural networks (BNN), BEAM, χ2 test (CHI) with 2 d.f., gradient boosted trees (GBM), and MDR.  The results are stratified by minor allele frequency (MAF).}
\label{fig:10nomargin}
\end{figure}
%\end{lscape}

BNN outperform all methods from the previous section (BEAM and $\chi^2$ test) by a very wide margin. This suggests that BEAM may be less robust to detect causal SNPs in the absence of marginal effects than previously thought, as it never achieves 25\% power in any of the scenarios tested. Again, we find these results encouraging as they indicate that BNNs are indeed powerful relative to existing approaches. Additionally, BNN outperformed the GBM method in all but 2 scenarios, indicating that BNN maybe be more adept at detecting purely epistatic signals across a broad array of MAFs and effect sizes. MDR performs well across every parameter combination tested, but as mentioned previously it is incapable of performing this analysis on a GWAS scale due to the exhaustive search technique and the need to perform permutation testing to assess statistical significance. To conclude this section, we note that BNN was the only method that did well across a variety of genetic models, number of SNPs, and MAFs while being capable of scaling to GWAS-sized data. This provides evidence that BNN framework is deserving of further investigation as an analysis technique for association studies. 

\subsection{Sensitivity and Specificity Analysis of the ARD Test}
\begin{figure}[!h]
\centering
\includegraphics[width=\textwidth]{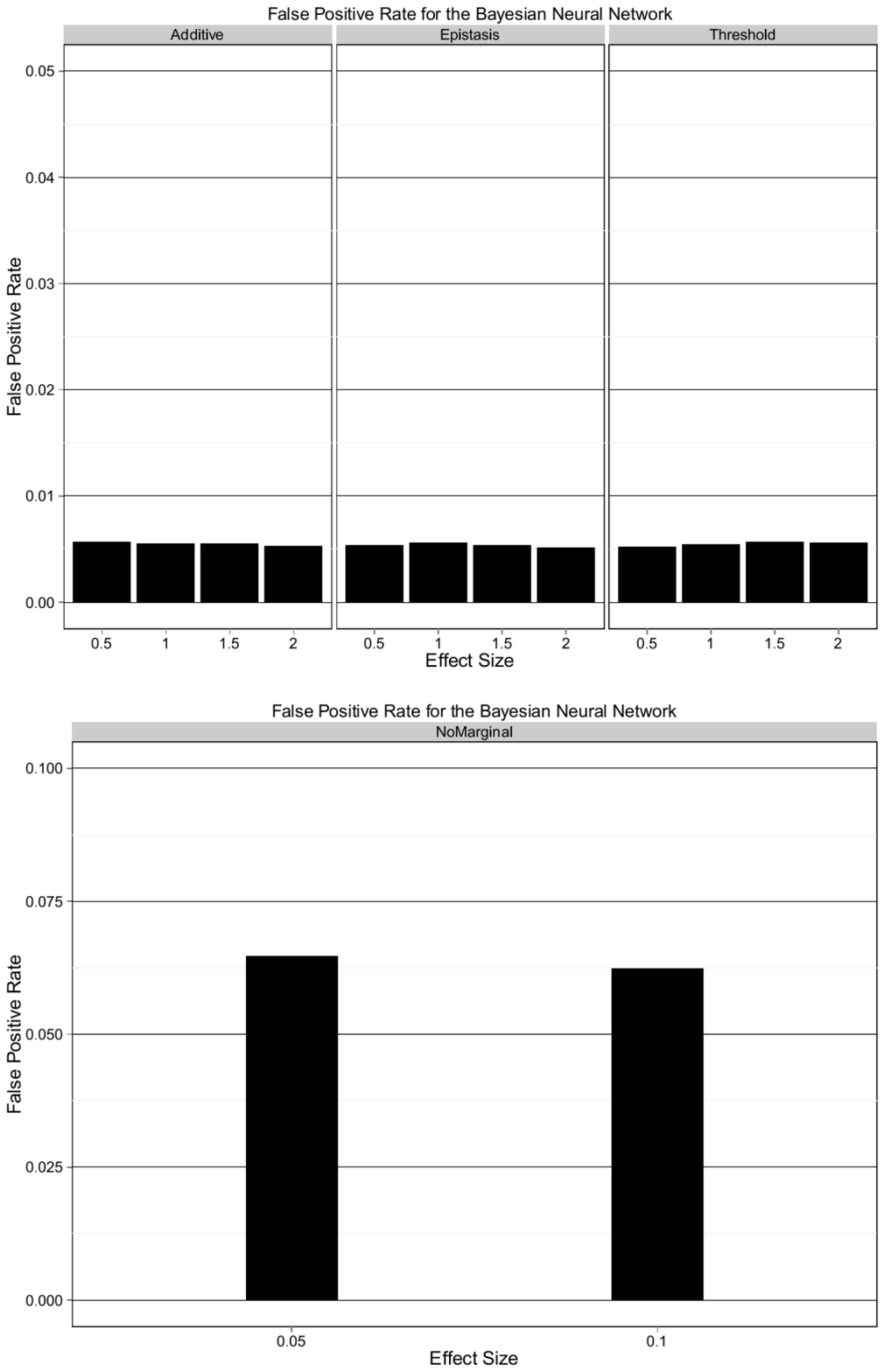}
\caption{False Positive Rates (FPR) for each model/effect size combination, averaged over MAF.}
\label{fig:fpr}
\end{figure}

The cutoff value used for the ARD test has an obvious impact on the method’s performance. In the extreme case, a cutoff of 0 would result in nothing being significant while a cutoff value of 1 would result in everything being declared as such. The cutoff value controls the tradeoff between sensitivity (i.e. the true positive rate) and specificity (i.e. the true negative rate, which is equivalent to 1 – the false positive rate). Evaluation of the false positive rate for the cutoff value of 0.6 used in the previous experiments indicates that the BNN method properly controls the amount of false positives. We observed an average false positive rate (FPR) of roughly 0.005 and 0.06 for the parametric models and the purely epistatic models, respectively as shown in Figure \ref{fig:fpr}. 

\begin{figure}[!h]
\centering
\includegraphics[width=\textwidth]{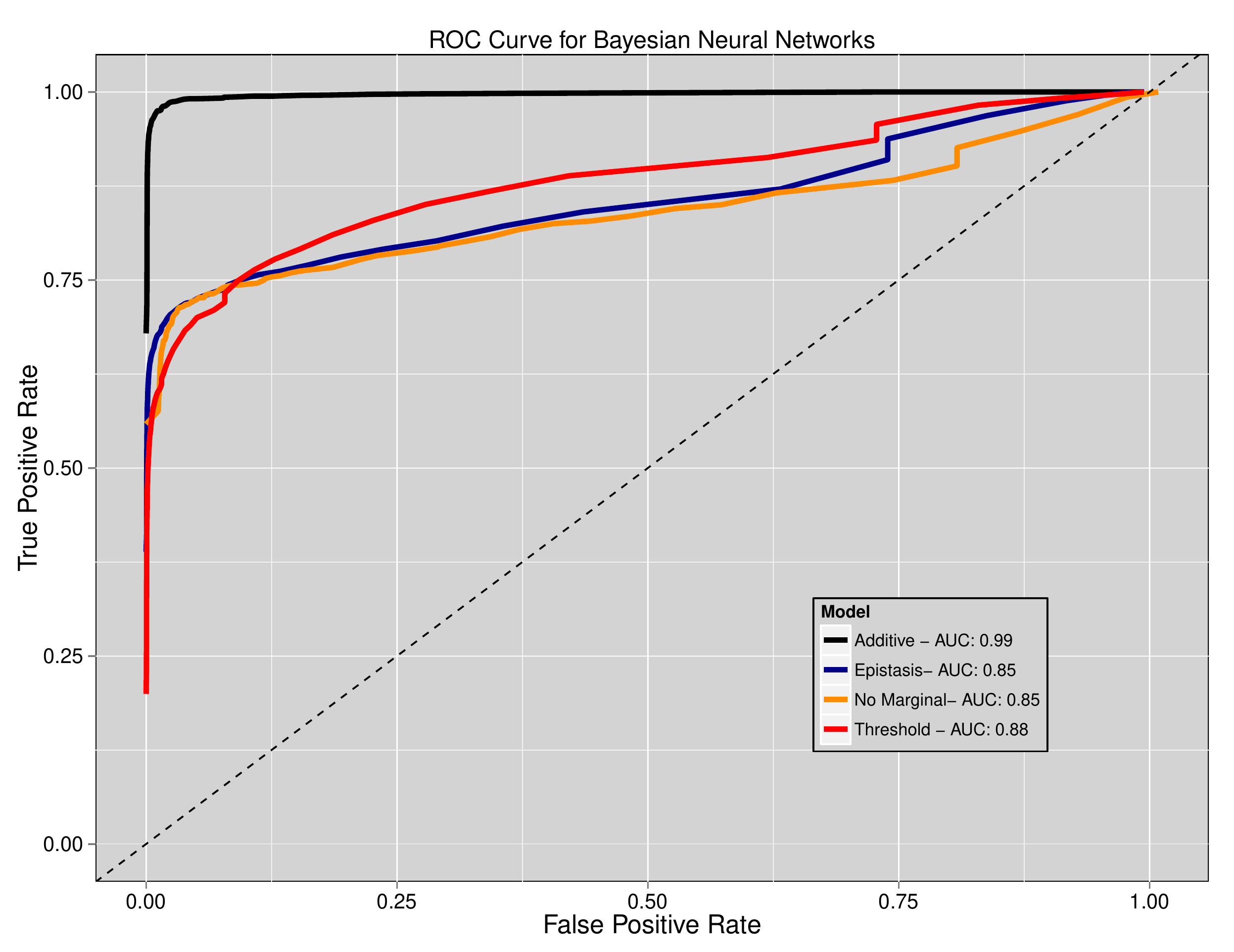}
\caption{Receiver-Operator Characteristic (ROC) curve for BNNs. Each line represents the ROC curve for a different genetic model, averaged over effect size and MAF. The area under the curve (AUC) for each model is shown in the legend.}
\label{fig:roc}
\end{figure}

To examine the trade off between the true positive rate (TPR) and FPR as the cutoff value is changed, we modulated the cutoff from 0 to 1 in increments of 0.01 and recorded the true positive and false positive rate for each data set in the two previous sections. In Figure \ref{fig:roc}, we averaged the TPR and FPR over effect size and MAF to produce a receiver-operator characteristic (ROC) curve for each of the 4 genetic models. The legend displays the area under the curve (AUC) for each model. 

These results show that BNN-ARD test for variable importance is able to achieve a high true positive rate, while maintaining a low false positive rate, which is an indication the method is performing as well and as expected. 

\subsection{Analysis of Tuberculosis Data}
To evaluate the performance of Bayesian neural networks on a real dataset, we analyzed a GWAS designed to find genetic markers associated with tuberculosis (TB) disease progression. The dataset describe in \cite{RefWorks:119}, contains information on roughly 60,000 SNPS and 105 subjects. For our study, each subject was classified as currently infected with any form of tuberculosis (i.e. extrapulmonary or pulmonary) or having a latent form of TB confirmed through a positive tuberculin skin test (purified protein derivative positive). Quality control was performed and SNPs with missing values were excluded, as were SNPs that were found to be out of Hardy-Weinberg equilibrium at the 0.05 level. After QC, there were 16,925 SNPs available for analysis and 104 subjects. Based on evidence of subpopulations in this data \cite{RefWorks:119}, subjects were assigned to one of three clusters created using the top two principal components and cluster membership was included as a covariate in the model.  Sampling of the Bayesian neural network was conducted as outlined in the previous section, with ARD hyper-parameters of $\alpha_0=3$, $\beta_0=1$. We performed 100 burn-in iterations followed by 1,000 sampling iterations which took approximately 20 hours. The top five SNPs based on posterior ARD probabilities are shown below in Table \ref{tab:tuber}.
\begin{table}[!h]
\caption{Top 5 SNPs based on posterior ARD probabilities. Note these probabilities are presented in terms of involvement (larger indicates a SNP is more likely to be involved).}
\begin{center}
\begin{tabular}{@{}ccc@{}}
\toprule
\textbf{SNP} & \textbf{CHR} & $\boldsymbol{Pr(\mu_j>\mu_{null})}$  \\ \midrule
rs966414           &      2        &            0.524                    \\ \midrule
rs1378124            &     8         &                  0.515              \\ \midrule
rs9327930            &       5       &                     0.509            \\ \midrule
rs4721214 & 7 & 0.502 \\ \midrule
rs4721214 & 9 & 0.498 \\ \bottomrule
\end{tabular}
\label{tab:tuber}
\end{center}
\end{table}
The SNP reported as the 2nd most significant in \cite{RefWorks:119} (rs10490266) appeared in our analysis as the 31st most significant SNP. Only one of the SNPs in Table \ref{tab:tuber} is currently known to be located within a gene (rs1378124 - MATN2) according to dbSNP. Every SNP reported in Table \ref{tab:tuber} is located on a the same chromosome and within 10-50 MB of loci previously reported as having a statistically significant association with pulmonary tuberculosis susceptibility \cite{png2012genome} in an Indonesian population. The loci reported in  \cite{png2012genome} were unfortunately either not part of the original SNP library or removed during the QC process in this study. Due to the small sample size of this dataset, it is hard to say conclusively which of the SNPs reported here and in \cite{RefWorks:119} are most likely to replicate in a larger study. However, we present this analysis to demonstrate that the BNN framework is capable of analyzing data sets containing a high number of SNPs in a relatively short amount time. 

\section{Conclusions}
In this study we have proposed the use of Bayesian neural networks for association studies. This approach was shown to be powerful across a broad spectrum of different genetic architectures, effect sizes, and MAFs. Of the approaches that do not rely on permutation testing, BNN was uniformly more powerful than the standard χ2 test and almost uniformly more powerful than the powerful than the popular BEAM method in the scenarios considered. BNN again showed a near uniformly better performance than the GBM method. MDR was very competitive with BNN in our evaluations, however MDR is incapable of scaling to larger datasets due to both its exhaustive search technique and reliance on permutation testing. In conclusion, we have demonstrated that BNNs are a powerful technique for association studies while having the capability of scaling to large GWAS sized datasets.

\section*{Availability of Code}
Source code implementing the GPU-based Bayesian neural network framework outlined in this paper is available at \texttt{https://github.com/beamandrew/BNN.}

\bibliography{bnn_case_control} 
\bibliographystyle{apalike}

\end{document}